\documentstyle[aps,prl,epsf,floats]{revtex}
\begin{document}
\twocolumn[\hsize\textwidth\columnwidth\hsize\csname@twocolumnfalse\endcsname
\title{A large-${\cal N}$ analysis of the local quantum critical point
and the spin-liquid phase}
\author{S. Burdin$^{1,2}$, M. Grilli$^{2}$, and D.R. Grempel$^{3}$}
\address{
$^1$ Institut Laue-Langevin, B.P. 156, 38042 Grenoble Cedex 9, France
}
\address{
$^2$Dipartimento di Fisica, 
Universit\`a di Roma, ``La Sapienza'', and INFM Center for Statistical
Mechanics and Complexity, 
Piazzale A. Moro 2, 00185 Roma, Italy
}
\address{
$^3$ CEA-Saclay/DSM/DRECAM/SPCSI, 91191 Gif-sur-Yvette Cedex, France
}
\date{\today}
\maketitle
\widetext
\begin{abstract}
We study analytically the Kondo lattice model with an additional 
nearest-neighbor antiferromagnetic interaction in the framework of 
large-${\cal N}$ theory. 
We find that there is a local quantum critical point between two phases, 
a normal Fermi-liquid and a spin-liquid in which the spins are 
decoupled from the conduction electrons. The local spin susceptibility 
displays a power-law divergence throughout the spin liquid phase.  
We check the reliability of the large-${\cal N}$ results by solving by 
quantum Monte Carlo simulation the ${\cal N}=2$ spin-liquid problem 
with no conduction electrons and find qualitative agreement. We show
 that the spin-liquid phase is unstable at low temperatures, 
 suggestive of a first-order transition to an ordered phase.  

\end{abstract}
]
\narrowtext

The effects of the vicinity to a  quantum 
critical point (QCP) on the low-temperature physical properties 
of various systems are presently attracting great
attention~\cite{sachdev}. There is in 
particular growing consensus that the anomalous metallic
properties observed in some 
heavy-fermion materials as well as the presence of 
superconductivity near magnetic phases 
could be interpreted in terms of the proximity to a 
QCP~\cite{QCPhf}. The issue of 
QCPs in superconducting cuprates is also a hotly
debated subject~\cite{QCPhtsc}.

In the traditional picture of the QCP
in metallic systems~\cite{hertz,millis} collective 
spin-fluctuations of specific wavevectors become critical 
at the QCP. A substantially different scenario has recently been
suggested~\cite{sinature} in which the quantum phase
transition involves all wavevectors {\it simultaneously}. 
At this {\it local} QCP the local degrees of freedom develop 
long-range correlations in time~\cite{colemannature} 
leading to a divergent local 
susceptibility. At the onset of magnetism the energy scale
that characterizes the coherent metallic phase vanishes.

It was proposed~\cite{sinature}
 that this peculiar type of behavior arises in
 a particular 
Kondo-lattice model with nearest-neighbor magnetic RKKY interactions.
The model was studied within Extended Dynamical Mean-Field
Theory (EDMFT) which reduces the problem
 to an effective impurity model in which a
 localized spin interacts with fermionic and bosonic 
baths~\cite{smithsi,chitrakotliar}.
However, some important aspects of the self-consistent procedure 
used in Ref.~\onlinecite{sinature} to close the set of EDMFT 
equations require further scrutiny~\cite{colemannature}.

In this paper we use a large-${\cal N}$ approach to solve analytically
the Kondo-lattice-RKKY model of Ref.~\onlinecite{sinature} 
fully implementing the self-consistent 
EDMFT scheme. We supplement this large-${\cal N}$ analysis with
a Quantum Monte Carlo (QMC) investigation of the Kondo-decoupled
phase of interacting spins represented by the pure RKKY model.

We start from the following Hamiltonian, 
\begin{equation}
H=
\sum_{\langle i,j \rangle, \sigma} t_{ij}c_{i\sigma}^{\dagger}c_{j\sigma}
+
\tilde{J}_{K}\sum_{i}{\bf s}_{i}\cdot {\bf S}_{i}
+
\sum_{\langle i,j \rangle}J_{ij} {\bf S}_{i}\cdot {\bf S}_{j}\;.
\label{hamiltonian}
\end{equation}

The operators $c_{i\sigma}^{\dagger}$ and ${\bf S}_{i}$ represent
respectively  
the creation of a conduction electron of spin $\sigma=
\uparrow, \downarrow$ and a localized 
spin at the {\it i}th site of a regular lattice of size $N$. 
We denote by  ${\bf s}_{i}$ 
the local spin density of the conducting electrons. 
The hopping terms $t_{ij}$ correspond to a non interacting electronic
density of states 
 $\rho_{c}^{0}(\epsilon)=1/{\rm N} \sum_{k}\delta(\epsilon-\epsilon_{k})$  
and the antiferromagnetic RKKY couplings 
$J_{ij}$ are described
by a spectral density $\rho_{I}(\epsilon)=1/{\rm N}
\sum_{q}\delta(\epsilon-J_{q})$ where 
$\epsilon_{k}$ and $J_{q}$ are the Fourier transforms of the 
nearest neighbors couplings $t_{ij}$ and $J_{ij}$, respectively. We 
assume for simplicity that $\rho_{c}^{0}$ and $\rho_ {I}$ are even
functions  that 
vanish outside the intervals  $[-D,D]$ and $[-J,J]$, respectively.

In the EDMFT approach the above model is implemented
on a lattice with a large coordination number $z$, so that
both the hopping and the magnetic RKKY coupling are scaled as
$t_{ij}\to t_{ij}/\sqrt{z}, \, J_{ij}\to J_{ij}/\sqrt{z}$.

To leading order in a large-$z$ expansion 
we obtain the following 
effective action for the local degrees of freedom on the 
0-th site (say)~\cite{BGG} :
\begin{eqnarray}
{\cal A}
&=&
-\int_{0}^{\beta}\int_{0}^{\beta}d\tau d\tau'
\sum_{\sigma=\uparrow, \downarrow}
c_{\sigma}^{\dagger}(\tau)
{\cal G}_{0}^{-1}(\tau-\tau')
c_{\sigma}(\tau') \nonumber \\
&+&
{\tilde J}_{K}\int_{0}^{\beta}d\tau 
{\bf s} (\tau)\cdot {\bf S}(\tau) \nonumber \\
&-& 
\int_{0}^{\beta}\int_{0}^{\beta}d\tau d\tau'
{\tilde \chi}_{0}^{-1}(\tau-\tau')
{\bf S} (\tau)\cdot {\bf S}(\tau')\;,
\label{action}
\end{eqnarray}
where the correlators ${\cal G}_{0}^{-1}(\tau)$ and ${\tilde
\chi}_{0}^{-1}(\tau)$ are self-consistently determined 
 cavity fields~\cite{reviewDMFT,smithsi,chitrakotliar}.

There are two steps in the implementation of the EDMFT 
procedure~\cite{reviewDMFT,smithsi,chitrakotliar}. 
The first (and usually most difficult)
one consists in solving the local impurity problem for fixed  
${\cal G}_{0}^{-1}$ and $\tilde{\chi}_0^{-1}$ in order to find the 
spin-symmetric 
local electronic Green function 
$G_{c}(\tau)=-\langle T c_{\sigma}(\tau)c_{\sigma}^{\dagger}(0)\rangle$ 
and the local magnetic susceptibility 
$\tilde{\chi}_{loc}(\tau)=\langle T {\bf S}(\tau)\cdot {\bf
S}(0)\rangle$.
Next, the condition that the impurity site is equivalent to any
other lattice site is imposed through a set of  self-consistency relations
that allow us to express the cavity fields ${\cal G}_{0}^{-1}$ and
$\tilde{\chi}_0^{-1}$ in terms of  $G_c$ and $\tilde{\chi}_{loc}$.

Here we implement this procedure in the framework of a
 large-${\cal N}$ expansion~\cite{largeN}. 
The SU(2) spins are replaced by
SU(${\cal N}$) operators in the fermionic representation
$S_{\sigma,\sigma'}=f^\dagger_\sigma f_{\sigma'}-\delta_{\sigma\sigma'}/2$
with $\sigma,\sigma'=1,...,{\cal N}$. The $f$-fermions are
subject to the constraint
$\sum_{\sigma=1}^{\cal N} f^\dagger_{\sigma} f_{\sigma} = {\cal N}/2$,
 enforced through the introduction of a Lagrange-multiplier 
$i\lambda(\tau)$~\cite{largeN}. The coupling constants
 must now be rescaled so that physical
 quantities remain finite in the ${\cal N}\to\infty$ limit. 
A consistent rescaling  of all the couplings 
{\it in the Hamiltonian} (\ref{hamiltonian})
 is not possible without loosing the dynamical character
 of the spin fluctuations. We adopt instead a different approach 
in which we rescale the couplings {\it in the action} 
(\ref{action}) as ${\tilde J}_K \to J_K/{\cal N}$ 
and ${\tilde \chi}^{-1}_0 \to 
\chi^{-1}_0/{\cal N}$, and define the local susceptibility 
per spin component
$\chi_{loc}(\tau)={\cal N}^{-2}\sum_{\sigma\sigma'}
\langle f_{\sigma}^{\dagger}(\tau)f_{\sigma'}(\tau)
f_{\sigma'}^{\dagger}(0)f_{\sigma}(0)\rangle$. With this
rescaling all the 
terms in Eq.~(\ref{action}) are ${\cal O(N)}$. 

In the ${\cal N} \to \infty$ limit, the hybridization parameter 
$r = -J_K\langle c^{\dagger}_\sigma f_\sigma \rangle $, 
the Lagrange multiplier $\lambda$ and the chemical potential 
$\mu$ are determined from the saddle point conditions
 that may be written in the compact form
\begin{equation}
\left\{ -\frac{r}{J_{K}}, \frac{1}{2}, \frac{n_c}{2} \right\} 
= -\left\{ G_{fc}, G_{f}, G_{c}
\right\}(\tau=0^-)\;.
\label{colkondo}
\end{equation} 
$G_{f}(\tau)=-\langle Tf_\sigma(\tau) f^{\dagger}_\sigma (0)\rangle$,
$G_{c}(\tau)=-\langle T c_\sigma (\tau) c^{\dagger}_\sigma (0)\rangle$ and
$G_{fc}(\tau)=-\langle T f_\sigma(\tau)c^{\dagger}_\sigma (0)\rangle$
are the dressed 
$f$-fermion, conduction electron and mixed Green functions. The
effect of spin fluctuations is embodied in a self-energy correction 
$\Sigma_{f}(\tau)=-2\chi_{0}^{-1}(\tau)G_{f}(-\tau)$
to the bare $f$ propagator,
${\cal G}_{f}^{-1}(\omega_n)\equiv i\omega_n+\lambda-\Sigma_{f}
(\omega_n)$. The dressed Green functions are 
$G_{c}(\omega_n)=G_{c}^{0}(i \omega_n+\mu-r^2{\cal G}_{f}(\omega_n))$, 
$G_{f}(\omega_n)={\cal G}_{f}(\omega_n)+r^{2}{\cal G}_{f}^{2}(\omega_n)
G_{c}(\omega_n)$ and $G_{fc}(\omega_n)=r{\cal G}_{f}(\omega_n)G_{c}
(\omega_n)$, respectively. 
The usually difficult step
of obtaining $\chi_{loc}$ from the bath $\chi_0^{-1}$ can 
easily be performed in the large-${\cal N}$ limit in which
\begin{equation}
\chi_{loc}(\tau) = -G_{f}(\tau )G_{f}(-\tau ).
\label{chiloclargeN}
\end{equation}

The parameter $r$ describes the binding of the
$c$ electrons to the $f$ (``spin'') degrees of freedom. If $r$ is
finite 
there is Kondo
compensation of the localized spin.
For sufficiently small values of $J$ we find 
a Fermi-liquid (FL) phase with $r \ne 0$ for temperatures $T\le T_K(J)$,
the Kondo temperature of the system. 
In this low-$T$ and low-$J$ region, the FL is characterized by a finite
coherence energy scale $\epsilon_{FL} \propto T_K$ and
a large Fermi surface containing $1+n_c$ states
(Luttinger theorem). 
The physical properties of this heavy Fermi liquid are similar to 
those discussed in Ref.~\cite{BGG}. In particular, the local 
magnetic susceptibility $\chi_{loc}(T=0, \omega=0)$ remains finite. 
Upon increasing $J$,
the physical quantities characterizing the Kondo-screened FL
($r$, $T_K$, and $\epsilon_{FL}$) gradually decrease and vanish
at a critical value of the magnetic coupling $J_c \sim T_K^0$, the
Kondo temperature for $J = 0$. 
At the same time
$\chi_{loc}(T=0, \omega=0)$ increases continuously and diverges at 
$J=J_c$. The paramagnetic solution with $r = 0$ can be described in terms 
of a gas of free $c$ electrons with a small Fermi surface enclosing $n_c$
states decoupled from a spin-liquid (SL) of strongly correlated $f$ fermions.
Notice that, since our large-${\cal N}$ approach 
neglects the fluctuations of $r$,
we cannot distinguish between the critical behavior {\it at} the QCP and
that inside the SL phase.

Of course, a global magnetic instability may occur at a finite temperature 
$T=T_c$ before this local instability of the Fermi surface has a
chance to 
develop. This will happen if $\chi(q=Q, \omega=0, T)$ diverges at $T
= T_c$, where $J_Q=-J$ corresponds to the
lower edge of the RKKY spectral density $\rho_I$.
In this case, the local QCP behavior will be  
hidden by the magnetic instability but its effects may still be
observable 
provided that spin ordering is established at sufficiently low 
temperatures. 

In order to proceed we must fix 
 the actual prescription used to
 close the self-consistency loop relating 
$\chi_0^{-1}$ to $\chi_{loc}$.
Following the EDMFT scheme of
 Refs.~\onlinecite{sinature,smithsi,chitrakotliar},
we introduce the local conduction electron and magnetic self-energies,  
$\Sigma_c(\omega_n)={\cal G}_{0}^{-1}(\omega_n)
-G_{c}^{-1}(\omega_n)$, and 
 ${M}(\omega_n)=\chi_0^{-1}(\omega_n)+
\chi_{loc}^{-1}(\omega_n)$. These quantities may be expressed 
 in terms of ${\cal O}({\cal N}^0)$ physical quantities as 
\begin{eqnarray}
G_{c}(\omega_n)
&=& \frac{1}{\rm N} 
\sum_k G_c(k,\omega_n)=
\int_{-\infty}^{+\infty}d\epsilon
\frac{\rho_{c}^{0}(\epsilon)}{i\omega_n-\Sigma_{c}(\omega_n)-\epsilon}\;,
\label{gc}\\
\chi_{loc}(\omega_n)
&=&
\frac{1}{\rm N} \sum_q \chi(q,\omega_n)=
\int_{-J}^{J}d\epsilon
\frac{\rho_{I}(\epsilon)}{M(\omega_n)+\epsilon}\;.
\label{chiloc}
\end{eqnarray}
with $\chi^{-1}(q,\omega_n)= M(\omega_n)+J_q $
and $G^{-1}_c(k,\omega_n)=  
\omega_n-\Sigma_{c}(\omega_n)-\epsilon_k $.

The results depend crucially on the behavior of $\rho_{I}(\epsilon)$
near the band edge. We discuss first the case in which 
the RKKY spectral density $\rho_I(\epsilon)$ 
is finite at the lower edge of the band, 
$\rho(-J)\equiv\rho_I$, a situation that arises when the spin
correlations have a two-dimensional
character~\cite{sinature,smithsi}. Then we may
explicitly integrate Eq.~(\ref{chiloc}) to find 
\begin{equation}
\chi_{loc}(\omega_n)=\rho_I \ln \left[\frac{M(\omega_n)+J}{M(\omega_n)-J}
\right]-{\cal R}(\omega_n)\;,
\label{chilogar}
\end{equation}
where ${\cal R}(\omega_n)$ is a regular function of frequency.
Notice that in this case
 a second-order magnetic transition [signaled by $M(\omega_n
= 0) = J$] is necessarily accompanied by a divergence of
$\chi_{loc}$. Such a divergence  at $T=0$
can only take place when $r$ vanishes,
that is at the quantum critical point, $J =  J_c$, and inside the
SL phase $J > J_c$.

A complete solution of the model at the QCP and in the SL phase
can be obtained in the large-${\cal N}$ approach 
by solving Eqs.~(\ref{colkondo}) and 
(\ref{chiloclargeN}) for $r=0$ supplemented with the relation 
\begin{equation}
\chi_0^{-1}(\omega_n)= J\coth \left[ 
\frac{\chi_{loc}(\omega_n) + {\cal R}(\omega_n)}{2\rho_I}
\right]-\frac{1}{\chi_{loc}(\omega_n)}\;,
\label{chi0solution}
\end{equation}
which follows easily from Eq.~(\ref{chilogar}).

Since the solution obtained from the RG treatment 
for the local impurity problem~\cite{sinature} yields a power-law 
behavior for $M$ and  a logarithmic divergence for $\chi_{loc}$
 at low $\omega$ and $T$, we first check 
the validity of these results within our large-${\cal N}$ scheme. 
We thus assume that, at low frequency, 
 $M(\omega_n) - J  \sim \left(\vert \omega_n\vert /J \right)^\gamma$ and 
$\chi_{loc}(\omega_n)=
 \gamma \rho_I\ln \left(J/ \vert \omega_n \vert
\right)$. Analytic continuation to real frequencies
 leads to:
\begin{equation}
{\rm Im}\left[\chi_0^{-1}(\omega) \right] =
\frac{\pi {\rm sgn}(\omega)}{2\gamma \rho_I\ln^2 \left | J/\omega \right |}.
\label{chi0realfreq}
\end{equation} 
The imaginary part of the self-energy $\Sigma_f(\omega)$ 
for real frequencies is 
\begin{eqnarray}
\Sigma''_f(\omega) &=& 
2\int_{-\infty}^{+\infty}d\varepsilon 
\;{\rm Im}\left[\chi_0^{-1}(\varepsilon) \right]\rho_{f}(\omega-\varepsilon)
\times \nonumber \\
&&[n_{B}(-\varepsilon)+n_{F}(\omega-\varepsilon)]\;.
\label{Imself}
\end{eqnarray}
In the low-frequency regime, with the (self-consistently verified)
assumption that $\Sigma_f(\omega)$
dominates $\omega$, the corresponding fermionic Green function
is given by $G_f(\omega)\approx -\Sigma_{f}^{-1}(\omega)$. 
Taking then $\rho_f(\omega)=
\sqrt{\gamma \rho_I}/ \left(2\pi \right)\ln |J/\omega|/
 \sqrt{\vert \omega \vert}$
[corresponding to $\Sigma''_f(\omega)= -\sqrt{\vert \omega\vert} /
\left( \sqrt{\gamma \rho_I}\ln \vert J/ \omega \vert \right) $], 
and with the correlator of Eq.~(\ref{chi0realfreq}),  
one can check that Eq.~(\ref{Imself}) is verified.
Using these results in Eq.~(\ref{chiloclargeN}) 
 we find $\chi_{loc}(\omega_n)= (\gamma \rho_I/6\pi)\left[
\ln \vert J/\omega_n \vert \right]^3$. We now notice that this finding
is inconsistent with the previously assumed simply logarithmic behavior
of $\chi_{loc}$. On the other hand, one can further check that
an {\it Ansatz} $\rho_f(\omega)\sim 1/\sqrt{\vert J \omega \vert}$
(necessary to get a purely logarithmic $\chi_{loc}$) does not
verify Eq.~(\ref{Imself}).
All the above shows that, within our large-${\cal N}$ scheme,
a correlator of the form $\chi_0^{-1}(\omega_n)\propto \ln^{-1}(J/\omega_n)$
produces additional logarithmic corrections to the local susceptibility,
which prevent self-consistency.

On the other hand, one can still find self-consistent solutions of
the problem.
Starting from the {\it Ansatz}
\begin{eqnarray}
\label{ansatzchiloc}  
\chi_{loc}(\omega_n)&=&(A/J)\left[ \vert\omega_n \vert/J
\right]^{-\delta}\;,\\ 
\label{ansatzgf}G_f(\omega_n) &=&
-i(B/J)\left[\vert\omega_n \vert/J \right]^{-\alpha}\;,
\end{eqnarray}
diverging at low frequencies, one finds ${\rm Im}\chi_0^{-1}(\omega)\approx 
(J/A)(\vert \omega\vert/J)^\delta
{\rm sgn}(\omega) \sin(\delta\pi/2)$. Inserting these expressions in
Eq.~(\ref{Imself}) one can determine ($T=0, \omega > 0)$
\begin{eqnarray}
\nonumber
\Sigma''_f(\omega)& =& 
-(2JB/\pi A)
\left(\omega/J\right)^{\delta-\alpha+1}\\
\label{sigmaout}
& \times &  \sin\left(\pi\delta/2\right)
\cos\left(\pi\alpha/2\right)
\int_{0}^{1}dx x^\delta/(1-x)^{\alpha}\;.
\end{eqnarray} 
Assuming that $\delta <\alpha$ (that will be self-consistently verified), 
one has $G_f(\omega)\approx -\Sigma_{f}^{-1}(\omega)$. Compatibility
between  
Eqs.~(\ref{ansatzgf}) and (\ref{sigmaout}) requires
\begin{eqnarray}
\label{exponent}
2\alpha & =&  \delta +1\;,\\ 
\label{factor1}
\frac{\pi A}{B^2}  
&=& 2 \sin\left(\frac{\pi\delta}{2}\right) 
\int_{0}^{1}dx \frac{x^\delta}{(1-x)^\alpha}\;.
\end{eqnarray}
On the other hand, self-consistency requires that $\chi_{loc}$ 
calculated from Eq.~(\ref{chiloclargeN}) for the specific
 $G_f$ considered here coincides with Eq.~(\ref{ansatzchiloc}). This
leads to the additional condition 
\begin{eqnarray}
\frac{\pi A}{B^2} \sin\left(\frac{\pi\delta}{2}\right) 
& = & \cos^2\left(\frac{\pi\alpha}{2}\right) 
\int_{0}^{1}\frac{dx}{x^\alpha (1-x)^\alpha}\\
& = &2 \sin^2\left(\frac{\pi\delta}{2}\right) 
\int_{0}^{1}dx \frac{x^\delta}{(1-x)^\alpha}\;,
\label{factor2}
\end{eqnarray}
where the last equality follows from Eq.~(\ref{factor1}).
Surprisingly, this equation (in conjunction with
Eq.~(\ref{exponent})) can be 
solved analytically. We find $\delta=2\alpha-1=1/3$. 
\begin{figure}[t]
\label{fig.1}
\epsfxsize=3in
\epsffile{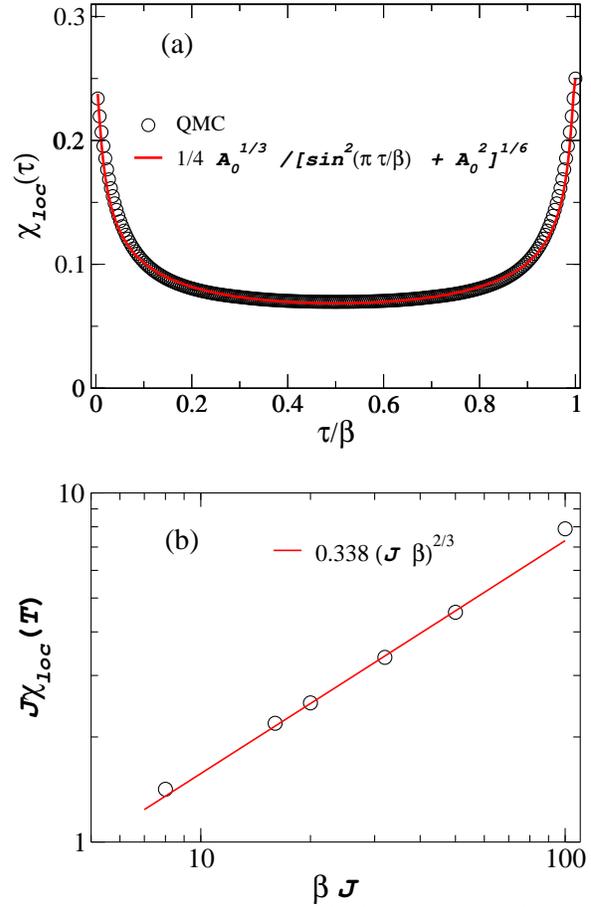}
\caption{Quantum Monte Carlo results for the decoupled model. 
We used up to 512 Trotter time-slices and    
$1.5 \times 10^5$ MC steps/time-slice. (a): 
$J\chi_{loc}(\tau)$ for $J/T = 50$. 
$A_0\approx 0.02059$ is a short-time cutoff 
obtained fitting the data to
the functional form shown in the legend. (b): $T$-dependence
of the static local susceptibility}
\end{figure}
We have tested these large-${\cal N}$ findings by a 
numerical approach based on the QMC algorithm of
Ref.~\onlinecite{daniel-marcelo}.
Since no algorithm exists capable of treating the Kondo and RKKY 
interactions on the same footing, we
confine ourselves to the investigation
of the paramagnetic Kondo-decoupled SL phase. Therefore we solve the 
self-consistency equations  for the pure RKKY model represented by
the last term of the action Eq. (\ref{action})
for standard SU(2) spins with a constant
$\rho_I(\epsilon)=1/(2 J)$. The local susceptibility in imaginary time
thus obtained is shown in Fig. 1(a). The low-$T$, long-time behavior
deduced from the numerical results is $\chi_{loc}(\tau) \sim [\sin(\pi
\tau/\beta)]^{-1/3}$. This corresponds to the following scaling form
\begin{equation}
\label{imchiloc}
\chi''_{loc}(\omega)\sim J^{-1}(\omega/J)^{-\delta} {\cal F}_\delta(\omega/T)
\end{equation}
with ${\cal F}_\delta(x)=x^{\delta}\vert \Gamma({1-\delta \over 2} 
+i{x \over 2\pi})
\vert^2 \sinh\left({x \over 2} \right)$ and $\delta = 2/3$.
Eq.~(\ref{imchiloc}) gives a power-law divergence at low energy, 
$\chi''_{loc}(T=0,\omega) \sim J^{-\frac{1}{3}}\omega^{-{2\over 3}}$ and 
$\chi'_{loc}(T,\omega=0)\sim J^{-\frac{1}{3}}T^{-{2\over 3}}$. 
The $T$-dependence of $\chi_{loc}(\omega = 0)$, displayed  
in Fig.~1(b), shows indeed this behavior. Quite interestingly, apart from
the difference in the value of the exponent 
$\delta$ (that was expected as critical exponents do 
depend on ${\cal N} $) the ``exact'' QMC results for SU(2) spins 
are qualitatively similar to those of the simpler large-${\cal N}$ approach.  
We expect on general grounds that the inclusions of $1/{\cal N}$ corrections 
should improve the agreement between the $1/{\cal N}$ expansion and QMC 
results. It is important to notice that if the solution 
(\ref{imchiloc}) for $\chi_{loc}$ is used  to determine 
$M(\omega)$ from Eq.~(\ref{chi0solution}), the resulting 
$M''(\omega)$ exhibits oscillations below a temperature
$T^{\star}$, a behavior that violates the condition of thermodynamic
stability $\omega M''(\omega) \le 0$. We have determined $T^{\star}$ 
in two different ways, from our analytic expressions
in Eqs. (\ref{chi0solution}) and (\ref{imchiloc}), and by 
analytically continuing our numerical results on the imaginary 
frequency axis using the method of Pad\`e approximants. The two methods yield 
$T^*\approx 0.03J$. This is proportional to, but substantially 
smaller than $J$, so that a wide range  $T^*<T<<J$ of temperatures 
still  exists within which the power-law given by~Eq.(\ref{imchiloc}) 
holds. The instability found at $T^*$ could signal
that the magnetic transition is first-order as
suggested by other authors~\cite{gabi} (see also Ref. \cite{rosch}); 
we will address this issue elsewhere~\cite{collaboration}.

We discuss next the case in which $\rho_I(\epsilon)$ has a square root
singularity at the band edge which corresponds to a three 
dimensional spin-wave 
spectral density. A particularly simple case is
that of the semi-circular density of states, 
$\rho_I(\epsilon)=(J\pi/2)^{-1}\sqrt{1-(\epsilon/J)^2}$. 
Then it is easy to show that 
$\chi_0^{-1}(\omega_n)=(J^2/4 )\chi_{loc}(\omega_n)$~\cite{reviewDMFT}. 
This particular model was studied in Ref.~\cite{BGG} in the context of  
the SL state 
first discussed in Ref.~\cite{sachdevye}. 
At the local QCP, the fermionic Green function diverges 
like $G_f(\omega)\sim 1/\sqrt{J\vert\omega\vert }$ which yields
$\chi_{loc}(\omega) \sim J^{-1}\ln(J/|\omega|)$ allowing to close 
the self-consistency
procedure {\it without} the appearance of additional logarithmic corrections. 
These results  hold as long as the magnetic self-energy 
$M(\omega_n)$ remains outside the  interval $[-J,J]$, 
so that no magnetic instability occurs. However, 
in this case $M(\omega_n =0) = J$ 
for $J \chi_{loc}(\omega_n = 0) = 2$, a condition satisfied
at a finite temperature $T_c \sim J$. 
Since at the transition $J < T^0_K$, it occurs while $T_K$ (and
$r$) are still finite. 
Then, as pointed out in Ref.\onlinecite{sinature}, 
in the three dimensional 
case the local QCP is
 masked by the finite temperature magnetic transition. 
The analysis of this QCP, expected
 to be of the Millis-Hertz type~\cite{hertz,millis}, lies beyond the
power of our lowest-order large-${\cal N}$ theory.    

In summary, we carried out a large-${\cal N}$ implementation of the 
EDMFT equations for the Kondo-lattice-RKKY model. This simple scheme 
allows us to find self-consistent solutions of the EDMFT equations
without any {\it ad hoc} input. We find in agreement with Ref.~\cite{sinature}
 that there exists a local QCP separating a FL and a SL phase. However, 
this approximation, which neglects the fluctuations of $r$,
does not allow us to distinguish between the properties of the QCP 
and those of the SL phase which correspond respectively to the 
unstable and stable fixed points
of the RG equations of Ref. \cite{sinature}. 
We showed that the large-${\cal N}$ results for the SL phase are reliable 
by comparing them with those of an exact numerical solution of the 
SL problem for  ${\cal N}=2$. Both our methods 
(in agreement with the RG results of Ref. \cite{siPRB}) show 
that a correlator $\chi_0^{-1}\sim \omega^\delta$ corresponds to   
a local susceptibility $\chi_{\rm loc} \sim \omega^{-\delta}$. 
On the other hand we find that
a correlator $\chi_0^{-1}\sim \ln^{-1}\omega$ does {\it not} 
correspond to a simply logarithmic susceptibility, showing
 that the limit $\delta \to 0$ can not be taken trivially. 
Although this conclusion can only 
 be safely drawn in connection with the SL problem, 
it suggests that caution must be 
exerted when taking the same limit at the QCP \cite{notascaling}. 

Finally, we showed that the power-law solution that we found in the SL phase
 is thermodynamically unstable below a temperature $T^{\star} \ll J$. 
This suggests that a first-order phase transition could take place at 
low temperature as suggested by others~\cite{gabi}.

We thank A. Georges for discussions and stimulating remarks. We also thank 
C. Castellani, A. Chubukov, B. Coqblin, C. Di Castro, C. Lacroix, 
Q. Si and A. Tsvelik for useful discussions.
One of us (S.B.) acknowledges  financial support from
the FERLIN Program of the European Science Fundation.

\vspace {-0.5 truecm}


\begin{references}
\vspace {-1. truecm}
\bibitem{sachdev} S. Sachdev, {\it Quantum Phase Transitions}, 
	Cambridge University Press, Cambridge (1999).
\bibitem{QCPhf} See, e.g., P. Coleman, Physica B {\bf 259-261}, 353 (1999).
\bibitem{QCPhtsc} A. Chubukov, {\it et al.}, Phys. Rev. B {\bf 49}, 
11919 (1994); C. Castellani, {et al.}, Phys. Rev. Lett. {\bf 75}, 4650 (1995);
C. M. Varma, {\it ibid.} {\bf 83}, 3538 (1999) and references therein; 
M. Vojta {\it et al., ibid.} {\bf 85}, 4940 (2000). 
\bibitem{hertz} J. A. Hertz, Phys. Rev. B {\bf 14}, 1165 (1976).
\bibitem{millis} A. J. Millis, Phys. Rev. B {\bf 48}, 7183 (1993).
\bibitem{sinature} Q. Si {\it et al.}, Nature {\bf 413}, 804 (2001);
	Q. Si {\it et al.} cond-mat/02024141.
\bibitem{siPRB}  L. Zhu and Q. Si, Phys. Rev. B {\bf 66}, 024426 (2002).
\bibitem{colemannature} For a general discussion of this issue
see, e.g., P. Coleman, Nature {\bf 413}, 788 (2001).
\bibitem{reviewDMFT} A. Georges {\it et al.}, 
Rev. Mod. Phys. {\bf 68}, 13 (1996).
\bibitem{smithsi} J. L. Smith and Q. Si, Phys. Rev. B {\bf 61}.
	5184 (2000).
\bibitem{chitrakotliar} R. Chitra and G. Kotliar, Phys. Rev. Lett.
	{\bf 84}, 3678 (2000). 
\bibitem{largeN}See, e.g., P. Coleman, Phys. Rev. B {\bf 35}, 5072 (1987); 
	D. M. Newns and N. Read, Adv. Phys. {\bf 36}, 799 (1987).
\bibitem{BGG} S. Burdin, D. R. Grempel, and A. Georges,
Phys. Rev. B {\bf 66}, 045111 (2002).
\bibitem{daniel-marcelo} D. R. Grempel and M.J. Rozenberg, 
Phys. Rev. Lett. {\bf 80}, 389, (1998).
\bibitem{gabi} S. Pankov, G. Kotliar, and Y. Motome, cond-mat/0112083. 
\bibitem{rosch} K. Haule, {\it et al.} cond-mat/0205347.
\bibitem{collaboration}S. Burdin {\it et al.},  unpublished.
\bibitem{sachdevye} S. Sachdev and J. Ye, Phys. Rev. Lett. {\bf 70},
	3339 (1993).
\bibitem{notascaling} Notice, however, that usually scaling properties
	holding at a critical point prevent the scaling quantities
	from acquiring (logarithmic) corrections except at the upper or 
	lower critical dimensions (for an example see, e.g., Appendix D, 
	in the second Ref.\cite{sinature}). 

\end{references}
\end{document}